\documentstyle[aps,pra,multicol,epsfig]{revtex}

\begin{document}

\draft

\title{Quantum Zeno-like effect due 
to competing decoherence mechanisms}

\author{Stefano Mancini and Rodolfo Bonifacio}

\address{
INFM, Dipartimento di Fisica,
Universit\`a di Milano,
Via Celoria 16, I-20133 Milano, Italy
}


\maketitle

\widetext

\begin{abstract}
We propose a selfconsistent quantum mechanical approach               
to study the dynamics of a two-level system
subject to random time evolution.
This randomness gives rise to competing effects
between dissipative and non-dissipative decoherence
with a consequent slow down of the atomic decay rate. 
\end{abstract}


\begin{multicols}{2}

\section{Introduction}

Time in Quantum Mechanics plays a rather ambiguous role.
Usually, it enters as a continuous parameter, but the principle
of relativity suggests a parallelism between position-momentum 
and time-energy. Thus, there was the attempt 
to quantize the time \cite{AB61}.
Otherwise, it has been considered as a discrete variable
\cite{B83,MIL91}. This approach implies a modification of the 
Schr\"odinger equation providing also an explanation of the 
nonappearance of macroscopically distinguishable states
in terms of non-dissipative decoherence.

More recently, a generalization of the Liouville-Von Neumann 
equation \cite{VON} 
was developed without any specific statistical 
assumption \cite{RB}.
It results an expression for the density operator  which, a 
posteriori, can be interpreted as if the evolution 
time is not a fixed 
parameter, but a stochastic variable whose distribution is 
a $\Gamma$-distribution.
It also provides a model for quantum measurement.
Furthermore, the characteristic time appearing in such approach
could be related to the time-energy uncertainty 
relation \cite{RB,MES}.
This approach is also useful to explain decoherence whenever the 
environmental degrees of freedom responsible for decoherence are 
not easily recognizable \cite{OLI}.

Here, we shall adopt this approach for a dissipative two-level 
system. Then, we shall show how decoherence effects compete
in this framework in different regimes.
In particular, we shall give an
example of how statistical quantum theory can describe 
a Zeno-like effect, i.e. frozen decay, which is
generally thought to arise 
from the state reduction 
caused by the measurements \cite{MISRA}.

\section{Standard approach}

Usually, the spontaneous decay 
of a two-level atom is described by 
means of a master equation derived by 
assuming the system interacting
with an environment \cite{EID}.

Consider a atom, with two relevant 
levels $\{|g\rangle\,,|e\rangle\}$
and lowering operator $\sigma=|g\rangle\langle e|$.
Let $\omega$ be the energy difference between the two levels
($\hbar=1$),
and let $\gamma$ be the decay rate.
Then, the master equation we are concerning is
derived by considering an interaction of the type 
$H\propto (\sigma\Gamma^{\dag}+\sigma^{\dag}\Gamma)$
where $\Gamma$ is a bath operator.
Under the Born-Markov approximation one arrives to \cite{QOPT}
\begin{equation}
{\dot\rho}={\cal L}\rho\,,\label{MED}
\end{equation}
with the Liouvillian superoperator
\begin{equation}
{\cal L}\rho=-i\omega\left[\sigma_{z},\rho\right]
+\frac{\gamma}{2}\left(
2\sigma\rho\sigma^{\dag}-\sigma^{\dag}\sigma\rho
-\rho\sigma^{\dag}\sigma\right)\,,\label{LIU}
\end{equation}
where, we choose to define 
the inversion operator as 
$\sigma_{z}=\sigma^{\dag}\sigma-\sigma\sigma^{\dag}$,
and the quadrature operators as
$\sigma_{x}=\sigma+\sigma^{\dag}$ and 
$\sigma_{y}=i\sigma-i\sigma^{\dag}$.

Dephasing processes, if necessary, 
are introduced in the same way.
These processes do not change the 
population of the two-level atom but 
do cause a phase randomization  of the atomic dipole. 
Then, they can be modeled by 
considering the inversion $\sigma_{z}$
to be coupled to environment by
$H\propto \sigma_{z} (\Gamma+\Gamma^{\dag})$. 
The master equation (\ref{MED}) now becomes
\begin{eqnarray}\label{MEDND}
{\dot\rho}=-i\omega\left[
\sigma_{z},\rho\right]&+&
\frac{\gamma}{2}
\left(2 \sigma\rho\sigma^{\dag}
-\sigma^{\dag}\sigma\rho
-\rho\sigma^{\dag}\sigma\right)
\nonumber\\
&-&\kappa\left[
\sigma_{z},\left[\sigma_{z},\rho\right]\right]\,,
\end{eqnarray}
where $\kappa$ represents the phase decaying rate.

From the master Eq.(\ref{MEDND}) it is 
easy to derive the following dynamical
equations 
\begin{eqnarray}
{\rm Tr}\left\{ {\dot\rho} \sigma_{z} \right\}
&=&-\gamma \, {\rm Tr}\left\{ \rho \sigma_{z} \right\} 
-\gamma \,,\label{SZEQ1}
\\
{\rm Tr}\left\{ {\dot\rho} \sigma \right\}
&=& - \left[i \omega+
\left(\frac{\gamma}{2}+\kappa\right)
\right]\,
{\rm Tr}\left\{ \rho \sigma \right\} \,.\label{SEQ1}
\end{eqnarray}
The solutions read
\begin{eqnarray}
{\rm Tr}\left\{ \rho(t) \sigma_{z} \right\}&=&
{\rm Tr}\left\{ \rho(0) \sigma_{z} \right\} 
\exp\left(-\gamma t\right)
\nonumber\\
&&+\left[ \exp\left(-\gamma t\right)
-1\right]\,,\label{SZ1}
\\
{\rm Tr}\left\{ \rho(t) \sigma \right\}&=&
{\rm Tr}\left\{ \rho(0) \sigma \right\} 
\exp\left[-i\omega 
t-\left(\frac{\gamma}{2}+\kappa\right)t
\right]\,.\label{S1}
\end{eqnarray}
We may see that the equation of motion for 
the inversion is unchanged
with respect to the dissipative case,
but the polarization decay rate is increased 
above the spontaneous 
emission result.

Nevertheless, decoherence 
is not always necessarily due to the 
entanglement with an environment but it may be due, 
especially the non-dissipative one, 
to the fluctuations of some classical parameters or internal 
variable of the system.
Or it might have an ``intrinsic" character \cite{GIULINI}.
Hence, we shall present a more general approach to  
non-dissipative decoherence for a two-level system.

\section{Random time evolution}

Quantum mechanics is a statistical theory whose elements
are ensembles of quantum systems, or ensembles of measurements
on the same quantum system.
This led, long time ago, to the introduction of the density 
operator \cite{VON}.
Along this line, we cannot state a priori that time is 
uniquely determined within the ensemble. 
Rather, it would be more reasonable to give a 
statistical interpretation of the time variable too. 
Then, following Ref. \cite{RB},
the evolution of a system is averaged on 
a suitable probability 
distribution $P(t,t')$ where $t'$ represents all possible 
times within the ensemble.
Let $\rho(0)$ be the initial state, then
the evolved state would be 
\begin{equation}\label{RHOBAR}
{\overline\rho}(t)=\int^{\infty}_0\, dt' \, P(t,t')\, \rho(t')\,,
\end{equation}
where $\rho(t')=\exp\{-i{\cal L}t'\}\rho(0)$ 
is the solution of the 
Liouville-Von Neumann equation \cite{VON}.

One can write as well 
\begin{equation}\label{RHOBARV}
{\overline\rho}(t)={\cal V}(t)\rho(0)\,, 
\end{equation} 
where the superoperator ${\cal V}$ is given by
\begin{equation}\label{VP} 
{\cal V}(t)=\int^{\infty}_0\, dt' \, P(t,t')\,e^{-i{\cal L}t'}\,.  
\end{equation} 
In Ref. \cite{RB}, the
function $P(t,t')$ has been determined to satisfy 
the following conditions:
i) ${\overline\rho}(t)$
must be a density operator, i.e. it must be self-adjoint, 
positive-definite, and with unit-trace.
This leads to the condition that $P(t,t')$ must be 
non-negative and normalized, i.e. a probability
density in $t'$, so that Eq.(\ref{RHOBAR}) is a completely 
positive map; ii) ${\cal V}(t)$ satisfies
the semigroup property 
${\cal V}(t_1+t_2)={\cal V}(t_1){\cal V}(t_2)$, 
with $t_1, t_2 \ge 0$. These requirements
are satisfied by 
\begin{equation}\label{V} 
{\cal V}(t)=\frac{1}{(1+i{\cal L}\tau)^{t/\tau}}\,,  
\end{equation} 
and  
\begin{equation}\label{P} 
P(t,t')=\frac{1}{\tau}\frac{e^{-t'/\tau}}{\Gamma(t/\tau)}
\left(\frac{t'}{\tau}\right)^{(t/\tau)-1}\,, 
\end{equation} 
where the parameter $\tau$ naturally appears as a scaling time.
Notice that the evolution superoperator (\ref{V}) only depends 
on $t$, and parametrically on $\tau$ as
in ``non extensive" generalization of Liouville
equation \cite{HECTOR}. Indeed, $t'$ comes out when 
a statistical interpretation is employed.
Expression
(\ref{P}) is the so-called $\Gamma$-distribution function, 
well known in line theory \cite{GNE}. 
The meaning of the parameter $\tau$ can be
understood by considering the mean $\langle t'\rangle=t$, 
and the variance
$\langle t'^2\rangle-\langle t'\rangle^2=\tau t$. 
Hence, $\tau$ rules the strength of time fluctuations,
or, otherwise, the characteristic correlation time of 
fluctuations. 

When $\tau\to 0$, $P(t,t')\to \delta(t-t')$ so
that ${\overline\rho}(t)\equiv\rho(t)$ 
and ${\cal V}(t)=\exp\{-i{\cal L}t\}$ 
is the usual evolution.

\begin{figure}[t]
\centerline{\epsfig{figure=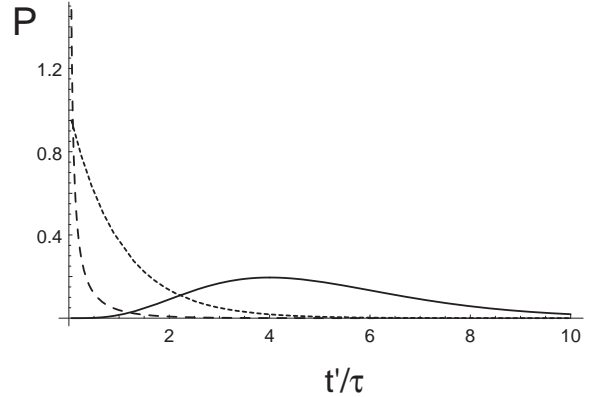,width=3.0in}}
\caption{\narrowtext 
Probability distribution (\ref{P}) as function of
dimensionless variable $t'/\tau$.
The dashed line refers to $t/\tau=0.1$, 
the dotted line to $t=\tau$,  
and the solid line to $t/\tau=5$.
}
\label{fig1}
\end{figure}

It is worth noting that
the behavior of the distribution (\ref{P}) strongly depends 
on the regime we consider. In fact, for $t\ll\tau$ we
have an exponential behavior, 
while for $t\gg\tau$ a Gausssian-like 
shape. The case $t=\tau$ 
represents the border between these two 
behaviors. All that is illustrated in Fig.(\ref{fig1}).

The phase diffusion
aspect of the present approach can also be seen in the 
evolution equation for the averaged density
matrix ${\overline\rho}(t)$. Indeed, by differentiating 
with respect to time Eq.(\ref{RHOBARV}) and
using (\ref{V}) one gets the following master equation 
for ${\overline\rho}(t)$
\begin{equation}\label{METOT} 
{\dot{\overline\rho}}(t)= -\frac{1}{\tau}\log
\left(1+i{\cal L}\tau\right) {\overline\rho}(t)\,. 
\end{equation} 

Once ${\cal L}\rho\equiv[H,\rho]$, 
the evolution operator ${\cal V}(t)$ 
describes a decay of the off diagonal
matrix elements in the energy representation, whereas the 
diagonal matrix elements remain constants,
i.e. the energy is still a constant of motion. In fact, 
in the energy eigenbasis,
Eqs.(\ref{RHOBARV}) and (\ref{V}) yield  
\begin{equation}\label{RHOBARNM}
{\overline\rho}_{n,m}(t)=
\exp\left(-\kappa_{n,m}t\right)\,
\exp\left(-i\nu_{n,m}t\right)
\,\rho_{n,m}(0)\,, 
\end{equation}
where 
\begin{eqnarray}
\kappa_{n,m}&=&\frac{1}{2\tau}
\log\left(1+\omega_{n,m}^2\tau^2\right)\,,\label{KANM}
\\
\nu_{n,m}&=&\frac{1}{\tau}
\arctan\left(\omega_{n,m}\tau\right)\,,\label{NUNM}
\end{eqnarray} 
with
$\omega_{n,m}$ the energy difference. 
One can recognize in Eq.(\ref{RHOBARNM}),
beside the exponential decay, a frequency shift of 
every oscillating term. 
This can be also used as a model for Quantum Nondemolition 
Measurement \cite{QOPT}.
In fact, in standard quantum 
measurement theory each measurement
results in an instantaneous 
reduction of the wave function onto
an eigenstate corresponding to 
the particular detected eigenvalue 
of the observable being measured.
Non-selective measurements
destroy the phase relation between 
different eigenstates and reduce the
state of the system to a statistical 
mixture where the non-diagonal 
elements of the corresponding density matrix vanish.
Therefore, all random dephasing events, 
i.e all processes that 
provide for a rapid quantum mechanical phase destruction
but leave the diagonal elements 
of the density matrix unchanged,
cause the same dynamical effect 
on the evolution of the system
like genuine quantum-nondemolition measurements.

However, it would also be possible to consider the Liouvillian 
(\ref{LIU}) in Eq.(\ref{METOT}), and 
therefore the competition between two types of decoherence. 
This is what we are going 
to study in the following.

\section{System dynamics with random time evolution}

If $\tau$ is small enough, one can expand
the logarithm in (\ref{METOT}) up to second order in $\tau$, 
and by using the Liouvillian (\ref{LIU}), we obtain 
\begin{eqnarray}\label{MEAPPROX}
{\dot{\overline\rho}}(t)&=&
-i\omega\left[\sigma_{z},{\overline\rho}(t)\right]
\nonumber\\
&&+\frac{\gamma}{2}\left(
2\sigma{\overline\rho}(t)\sigma^{\dag}
-\sigma^{\dag}\sigma{\overline\rho}(t)
-{\overline\rho}(t)\sigma^{\dag}\sigma\right)
\nonumber\\
&&-\frac{\tau}{2}\omega^{2}\left[\sigma_{z},
\left[\sigma_{z},{\overline\rho}(t)
\right]\right]\,,
\end{eqnarray}
where we have used $\omega\gg\gamma$ and
$\tau\ll\gamma^{-1}$, that is, the dissipation takes
place on a time scale much larger than the time fluctuations.

Eq.(\ref{MEAPPROX}) practically coincides with Eq.(\ref{MEDND})
provided to identify $\tau\omega^{2}/2$ with $\kappa$.
Nonetheless, the present approach is different from the
usual master equation approach, in the sense 
that it is model independent 
and without specific statistical assumptions.

\begin{figure}[t]
\centerline{\epsfig{figure=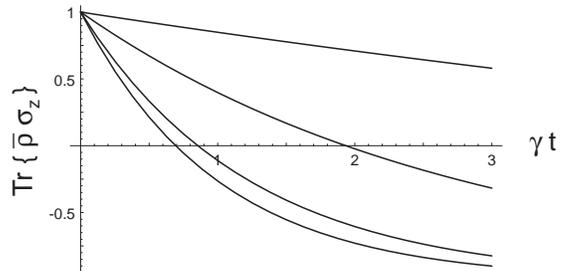,width=3.0in}}
\caption{\narrowtext 
Population inversion as function of 
the scaled (dimensionless) time $\gamma t$,
for different values of $\gamma\tau$.
From bottom to top $\gamma\tau=0$, $0.5$, $5$, $50$.
The system has been considered initially in the upper level.
}
\label{fig2}
\end{figure}

For a generic value of $\tau$, it is not possible to
extract an explicit form of master equation from Eq.(\ref{METOT}).
Nevertheless, the physics of the system can be understood 
by simply averaging the quantities of interest over the
distribution (\ref{P}). 
For instance, from Eqs.(\ref{METOT}) and (\ref{LIU}), we get
\begin{eqnarray}\label{SZ2}
{\rm Tr}\left\{{\overline\rho}(t) \, \sigma_{z}\right\}&=&
{\rm Tr}\left\{{\overline\rho}(0) \, \sigma_{z}\right\}
\exp\left[-\frac{t}{\tau}\log\left(1+\gamma\tau\right)\right]
\nonumber\\
&+&\left\{
\exp\left[-\frac{t}{\tau}\log\left(1+\gamma\tau\right)\right]
-1\right\}\,.
\end{eqnarray}
Equation (\ref{SZ2}) in the limit $\gamma\tau\ll 1$ 
reduces to the
usual decay described by Eq.(\ref{SZ1}). More generally,
the decay rate results modified. 
In particular, for $\gamma\tau\gg 1$
it would be possible to inhibit the dissipative effects through 
the nondissipative ones. 
The frozen dynamics due to increasing values of 
$\gamma\tau$ is shown in Fig.(\ref{fig2}).
This situation comes out as consequence of the transition
from the Gaussian to the exponential behavior of the
probability distribution (\ref{P}) (see Fig.\ref{fig1}).

The freezing effect on the system dynamics remind us 
the quantum Zeno effect \cite{MISRA}.
Its usual description rests on the suppression 
of the unitary Hamiltonian evolution of a quantum system due to 
intermittent measurements in rapid succession. 
Due to the wave function collapses, in the limit 
of continuos measurements, the evolution is completely inhibited,
and the system is frozen in its initial state.
Here, instead, the effect entirely arise 
from quantum statistical properties. 

While in the usual quantum Zeno effect 
the essential requirement is that 
the measurements of the system state,  
which cause the interruption,
be more closely spaced in time 
than the reservoir correlation
(memory) time, in our case the 
correlation time of fluctuations
should exceed the typical decay time.

Essentially, we may claim that in the limit $\gamma\tau\gg 1$ 
one type of decoherence prevents 
the other. In fact, we may think that 
dissipative decay process  
takes place through the energy channels 
determined by the system-environment 
interaction. However, the time evolution 
fluctuations make these channels completely fuzzy, 
thus preventing the decay.  

The above results can be easily extended 
to the case of driven 
two-level system.

\section{Conclusions}

In conclusion, we have presented a simple 
model able to explain different 
aspects of decoherence in a two-level system. 
The used formalism predicts, for large time fluctuations, 
a novel Zeno-like effect without invoking the abstruse 
concept of wavefunction collapse \cite{SCHE}.

The generality of the presented approach 
suggests in some way the 
possibility that the parameter $\tau$
(even though system-dependent)
might have a lower nonzero limit, 
related e.g. to the 
time energy uncertainty relation \cite{RB},
or to the finite extension of the 
spatial wavefunction \cite{JPHYSB},
or even to gravitational effects \cite{ELLIS}.
However, even if such 
``intrinsic" decoherence effects emerge, 
the value of $\tau$ would be 
very small. 
Nevertheless, one can think as well to 
introduce the above statistical properties 
by hand from the outside.
For instance, one can use dephasing processes
through a noisy driving field as
envisaged in Ref.\cite{KUR}.
Otherwise, one could think at a sequence of
measurements as jump-like processes, 
randomly distributed in time \cite{HERZ}.
In such a cases the statistics, 
hence the parameter $\tau$,
would be controlled by the experimenter.
Thus, it would be an interesting challenge
to arrange an experimental set up 
where the conditions for above 
Zeno-like effect are
achieved.

\section*{Acknowledgements}
We gratefully acknowledge useful discussions 
with David Vitali.

\end{multicols}


\begin{references}

\bibitem{AB61}
Y. Aharonov and D. Bohm, Phys. Rev. {\bf 122}, 1649 (1961).

\bibitem{B83}
R. Bonifacio, Lett. Nuovo Cimento 
{\bf 37}, 481 (1983).

\bibitem{MIL91}
G. J. Milburn, Phys. Rev. A {\bf 44}, 5401 (1991). 

\bibitem{VON}
J. von Neumann, 
{\it Mathematical Foundations of Quantum Mechanics},
(Princeton University Press, Princeton, 1955).

\bibitem{RB}
R. Bonifacio, Il Nuovo Cimento {\bf 114 B}, 473 (1999);
R. Bonifacio, in {\it Mysteries, 
Puzzles and Paradoxes in Quantum Mechanics},
Ed. by R. Bonifacio (AIP, Woodbury, 1999).

\bibitem{MES}
A. Messiah, {\it Quantum Mechanics}, 
(North-Holland, Amsterdam, 1961).

\bibitem{OLI}
R. Bonifacio, S. Olivares, P. Tombesi and D. Vitali,
Phys. Rev. A {\bf 61}, 053802 (2000).

\bibitem{MISRA}
B. Misra and E. C. G. Sudarshan, J. Math. Phys. {\bf 18}, 
756 (1977).

\bibitem{EID}
W. H. Zurek, Phys. Today {\bf 44}, 36 (1991).

\bibitem{QOPT}
D. F. Walls and G. J. Milburn, {\it Quantum Optics},
(Springer, Berlin, 1995).

\bibitem{GIULINI}
D. Giulini, E. Joos, C. Kiefer, J. Kupsch, 
I. O. Stamatescu and M. D. Zeh, 
{\it Decoherence and the Appearance of a Classical World
in Quantum Theory}, (Springer, Berlin, 1996).

\bibitem{HECTOR}
A. Vidiella-Barranco and H. M. Moya-Cessa,
Phys. Lett. A {\bf 279}, 56 (2001).

\bibitem{GNE}
See, e.g., B. V. Gnedenko, {\it The theory of probability},
(Chelsea, New York, 1962).

\bibitem{SCHE}
V. Frerichs and A. Schenzle,
Phys. Rev. A {\bf 44}, 1962 (1991).

\bibitem{JPHYSB}
S. Mancini and R. Bonifacio,
J. Phys. B: At. Mol. Opt. Phys. {\bf 34},
1909 (2001).

\bibitem{ELLIS}
J. Ellis, S. Mohanti and D. V. Nanopoulos,
Phys. Lett. B {\bf 235}, 305 (1990).

\bibitem{KUR}
G. Harel, A. G. Kofman, A. Kozhekin and G. Kurizki,
Opt. Express {\bf 2}, 355 (1998).

\bibitem{HERZ}
U. Herzog, Phys. Rev. A {\bf 52}, 602 (1995).

\end{references}
\end{document}